# Modeling and Simulation of Solar Photovoltaic Cell for the Generation of Electricity in UAE

Shadman Sakib
Department of Electrical and Electronic Engineering
International University of Business Agriculture and Technology
Dhaka 1230, Bangladesh
sakibshadman15@gmail.com

Md. Abu Bakr Siddique
Department of Electrical and Electronic Engineering
International University of Business Agriculture and Technology
Dhaka 1230, Bangladesh
absiddique@iubat.edu

*Abstract*—This paper proposes the implementation of a circuit based simulation for a Solar Photovoltaic (PV) cell in order to get the maximum power output. The model is established based on the mathematical model of the PV module. As the PV cell is used to determine the physical and electrical behavior of the cell corresponding to environmental factors such as temperature and solar irradiance, this paper evaluates thirty years solar irradiation data in United Arab Emirates (UAE), also analyzes the performance parameters of PV cell for several locations. Based on the Shockley diode equation, a solar PV module is presented. However, to analyze the performance parameters, Solarex MSX 120, a typical 120W module is selected. The mathematical model for the chosen module is executed in Matlab. The consequence of this paper reflects the effects of variation of solar irradiation on PV cell within UAE. Conclusively, this paper determines the convenient places for implementing the large scale solar PV modules within UAE.

*Keywords*—PV Module, Solar Irradiation, Maximum Power Point Tracking (MPPT), Temperature, Performance parameter, I-V and P-V characteristics, MSX 120PV Module

## I. INTRODUCTION

These days the whole world is facing many challenges to control the difficulty of the energy crisis. As energy is one of the basic needs for human activity, it acts as a prime mover for social as well as for economic development. But due to the rapid population growth and industrialization demands an increase in the amount of electrical energy hence renewable energy plays a key role to fulfill the power and energy crisis in today's world. The renewable energy sources which are also known as the non-conventional type of energy which includes solar energy, bio-energy, biofuels, biomass, wind energy, and tidal energy are most commonly utilized to meet the energy needs and considered as the fundamental solutions for energy crisis and environmental concerns [1, 2]. These renewable energy resources limit the emission of greenhouse gases and slows the depletion of fossil fuels [3]. Apart from the other renewable energy sources, the power generation from the solar energy using PV is one of the most promising renewable resources since it requires less maintenance, no wear, and tear, no direct pollution, reduces emissions, energy consumption produced by oil and gas, and involves no moving parts [4].

A solar PV system converts the sun radiation into electricity with the help of Photo Electric effect. A PV system consists of PV arrays and electric converters which converts the solar energy into Direct Current (DC) electricity when sunlight shines on the PV array. The DC power is converted into AC power through the inverter which is used to power the local loads or fed back to the utility services [5]. For all living being on earth, the sun is the primary source of almost all energy which supplies energy in the form of radiation. The massive amount of energy is produced by the nuclear fusion of hydrogen atoms inside the sun's core which compresses the hydrogen atom to turn into helium. The power from the sun seized by the earth is approximately $1.8 \times 10^{11}$MW [6]. The output characteristics of the PV module mainly depend on the solar insolation, the cell temperature, and the output voltage of the PV module. However, PV systems comprise of a PV generator, energy storage devices such as batteries, AC and DC inverter.

Maximum Power Point Tracking (MPPT) system is applied to the power converters to make sure the optimal use of the available solar energy [7]. It is essential to model MPPT for the design and simulation for the PV system applications as the PV module has nonlinear characteristics.

The United Arab Emirates (UAE) is located in the Middle East and in the eastern part of the Arabian promontory. It consists of seven emirates and has a land of area 83,600km$^2$ containing 9.68 million people in it and increasing day by day results in a significantly high amount of electricity demand each year. UAE contains 9.3 percent of the world's oil reserves and 4.1 percent of the world's gas reserves [8]. The use of energy along with its incorporated $CO_2$ emissions have been a controversial issue in the Gulf region over more than past one decade and UAE ranks 25[th] place for $CO_2$ emission. Electricity is the major form of energy consumed in UAE and it is produced through the use of fossil fuel which is mainly oil and gas. The generation of electricity has grown considerably also the consumption of fossil fuel as well as $CO_2$ emissions have increased enormously due to the rapid population growth and economic expenditure which is incorporated with massive architectural projects. For this reason, the UAE government started to invest in renewable energy technologies that will make this country less dependent on conventional energy. However, the importance of using renewable energy in UAE will meet the use of energy policies to ensure buildings sustainability and provide guidelines for future architecture [9] also contribute into the national grid to meet the peak-load demand during the summer seasons thereby reducing the power demands. Hence to estimate the average and maximum output power at different places it is important to study the solar radiation pattern and other climate situations.

In this paper, a mathematical simulation model for Solarex MSX 120, a typical 120W module is represented. A double diode model is chosen. The model is able to simulate both the I-V and P-V characteristics curve. Also, the model is used to study different parameters variations effects on the PV array which includes operating temperature and solar

## II. MODELING OF THE PHOTOVOLTAIC CELL

A solar cell is the building block of a solar panel as well as the fundamental unit of PV cell. A PV cell is created by the combination of many solar cells in series and parallel. Solar cells which are connected in series are used to increase the output voltage and the cells in parallel will produce a higher current. The electrical characteristics of the solar cell differ barely from a diode, the relationship between the cells terminal voltage and current is represented by the Shockley equation

$$I = I_s \left(e^{(V_d/nV_t)} - 1\right) \qquad (1)$$

Generally, a photovoltaic cell is a silicon semiconductor junction device composed of a p-n junction almost identical to a diode. It converts sunlight directly into electricity. When the P-N junction is visible to light, photons with energy greater than band gap energy of the semiconductor are absorbed creating the electron hole-pairs which are proportional to the incident irradiation. In the dark, the I-V output characteristics of a PV cell is similar to that of a diode [6]. When the cell is short circuited, this current flows in the external circuit; when open circuited, this current is shunted internally by the intrinsic P-N junction diode. The characteristics of this diode, therefore, set the open circuit voltage characteristics of the cell [10]. The equivalent circuit of a simple PV cell can be modeled by a current source in parallel with two diodes, a parallel resistor indicating a leakage current and a series resistor expressing an internal resistance to the current flow as outlined in figure 1 [11].

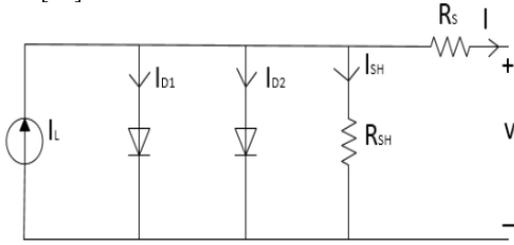

Fig. 1. Electrical equivalent circuit diagram of solar cell

The photocurrent $I_L$ induced by the PV cell is proportional to the solar insolation. The output current I of the cell is given by

$$I = I_L - I_{D1} - I_{D2} - I_{SH} \qquad (2)$$

For an ideal PV cell, there is no series loss as well as no leakage, i.e., $R_S = 0$ and $R_{SH} \approx \infty$ therefore, $I_{SH} \approx 0$. Therefore, the (2) can be rewritten as

$$I = I_L - I_{D1} - I_{D2} \qquad (3)$$

If $I_S$ is the saturation curent of the diode then the current that bypass through the diodes are given by

$$I_{D1,D2} = I_s\left[e^{q(V+IR_S)/AkT} - 1\right] \qquad (4)$$

Therefore, the current-voltage characteristic equation for the PV cell is given below

$$I = I_L - I_{S1}\left[e^{q(V+IR_S)/AkT} - 1\right] - I_{S2}\left[e^{q(V+IR_S)/AkT} - 1\right] \qquad (5)$$

Here, $I_{D1}$ and $I_{D2}$ are the currents passing through the individul diodes, $I_{S1}$ is the saturation current of the diode D1 and $I_{S2}$ is the saturation current of the diode D2, $q=1.6*10^{-19}$C is the charge of an electron, $k=1.38*10^{-23}$J/K is a Boltzmann's constant, V is the terminal voltage of the cell, $R_S$ is a series resistance, T is the working temperature and A is an ideal constant of the diode. The ideality constant varies depends on PV technology [12].

Photo generated current mainly influenced by the solar insolation and cell's working temperature, described by

$$I_L = \left[I_{SC} + K_I(T - T_r)\right]\beta \qquad (6)$$

$$I_L = \frac{G * I_{SC}(T_{1,nom})}{G_{(nom)}} \qquad (7)$$

Furthermore, the cell's saturation current $I_S$ fluctuates with the cell's working temperature, illustrated by the following equation

$$I_S = I_{RSC}(T/T_r)^3 * e^{\left[\frac{qE_{BG}}{kA}\left(\frac{1}{T_r} - \frac{1}{T}\right)\right]} \qquad (8)$$

$$I_S = I_{RSC} / \left(e^{qV_{OC}/AkT_r} - 1\right) \qquad (9)$$

Neglecting the parallel resistance $R_{SH}$ and considering the series resistance $R_S$, the resistance inside each cell in the connection between cells are given by

$$R_S = -dV/dI_{V_{OC}} - 1/X_V \qquad (10)$$

$$X_V = \left[I_S * q / nkT_r * \exp(qV_{OC}/AkT_r)\right] - 1/X_V \qquad (11)$$

Here, $I_{SC}$ is the cell's short circuit current at 1kWm$^{-2}$, $I_{RSC}$ is the cell's reverse saturation current, T is the cell temperature (K), G is the irradiation, $T_r$ is the cell's reference temperature, $E_{BG}$ is the band-gap energy, cell's, $R_S$ is the small resistance which expresses the internal losses, $V_{OC}$ is the open circuit voltage and β is the solar insolation in 1kWm$^{-2}$.

## III. CHARACTERISTICS OF PHOTOVOLTAIC CELL

Photovoltaic cells generally demonstrate a nonlinear I-V and P-V characteristics which vary with the solar irradiation and cell temperature. The most important fundamental parameters used for characterizing the photovoltaic cell are: short circuit current $I_{SC}$, open circuit volatge $V_{OC}$, Maximum Power Point (MPP), effiency (η) and Fill Factor (FF) [13].

<u>Short circuit current</u>: It is the current that reduces the effect of impedance in the circuit. When the cell is shortcircuited, negligible current flows in the diode. It is calculated when V=0. However, it is largest amount current produced from the PV cell due to the photon excitation.

$$I_{L(V=0)} = I_{SC} \qquad (12)$$

<u>Open circuit volatge</u>: It is voltage which is not connected to any load in a circuit and no current passing through the cell. It is calculated when the volatge is equal to zero. Moreover, it is the maximum voltage difference across the PV cell when I=0. Mathematically,

$$V_{OC} = \frac{AkT}{q}\ln(I_L/I_S + 1) = V_{th} * \ln(I_L/I_S + 1) \qquad (13)$$

Here, $V_{th}$ is the thermal volatge and T is operating temperature of the PV cell.

<u>Maximum Power Point</u>: It is the operating point where the power is maximum across the load. Mathematically,

$$P_m = V_m * I_m = \alpha V_{OC} * I_{SC} \qquad (14)$$

Here, α is the fill factor.

Efficiency: It is defined as the ratio of maximum power to the incident light power. Mathematically,

$$\eta = (P_m / P_{in}) * 100 = (V_m * I_m / P_{in}) * 100 \quad (15)$$

Fill Factor: Fill Factor which is abbreviated by FF, is a parameter defined as the ratio of the maximum power from the solar cell to the product of $I_{SC}$ and $V_{OC}$, expressed as

$$FF = P_m / V_{OC} * I_{SC} = V_m * I_m / V_{OC} * I_{SC} \quad (16)$$

Typically fill factors ranges from 0.5 to 0.82. Its value is more than 0.7 for good PV cells. The fill factor decreases with the increase of cell temperature.

In this experiment Solarex MSX120, a 120W PV module is used to to analyze the electrical performance parameters at different locations. The MSX120 module contains 72 multi-crystalline solar cells configured as 4 series srtings of 18 cells each. The model of the PV module was executed using a Matlab program. The program computes the current I using the electrical parameters of the module, variable voltage V, irradiance G and temperature T. PV module is influenced by temperature, light intensity etc. Newton Raphson method is used for solving cell current iteratively. The output power of PV cell and output current depends on cell's operating voltage, temperature and solar insolation. The key specification and electrical characteristics of MSX120, 120W PV cell is shown below in Table 1.

TABLE I. ELECTRICAL CHARACTERISTICS OF SOLAREX 120, 120W PV MODULE

| Parameter | Specification |
|---|---|
| Maximum Power ($P_m$) | 120W |
| Voltage at $P_m$, ($V_{mp}$) | 34.2V |
| Current at $P_m$, ($I_{mp}$) | 3.5A |
| Minimum $P_m$ | 114W |
| Short-circuit current ($I_{SC}$) | 3.8A |
| Open-circuit voltage ($V_{OC}$) | 42.6 |
| Temp. co-efficient of $I_{SC}$, ($K_I$) | (0.065±0.015)%°C |
| Temp. co-efficient of $V_{OC}$ | -(160±20)mV/°C |
| Temp. coefficient of power | -(0.5±0.05)%/°C |
| Maximum system voltage | 600V |
| Maximum series fuse rating | 20A |
| NOCT[2] | 47±2°C |
| Type of cell | multi-crystalline silicon |

## IV. AVERAGE SOLAR RADIATION IN UAE

Although UAE is an oil-reach country, but due to the increase in its population as well as huge architectural projects resulted in huge electricity demand in the country. To meet the high demand for electricity, it is producing power from renewable energy sources like solar and wind. Among them, solar energy is the most reliable and environmentally friendly renewable energy in UAE. Therefore, renewable energy sources play a key role in the development of alternative energy in UAE. According to NASA Surface Meteorology and Solar energy [14] website, Thirty years of average solar irradiation data is used to find out the variation of solar irradiation in UAE. The average solar irradiation in seven different emirates throughout spring (March-May), summer (June-August), autumn (September-November) and winter (December-February) are shown in figure 2, 3, 4 and 5.

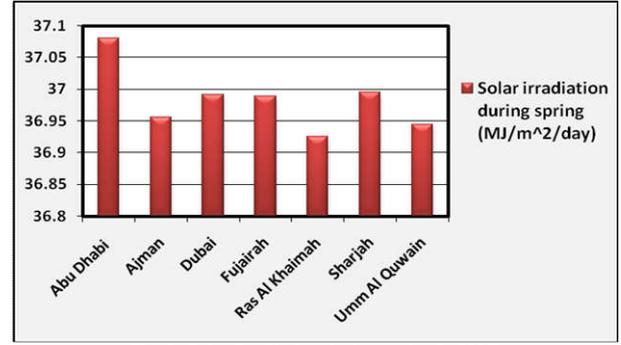

Fig. 2. Solar Irradiation during spring in UAE

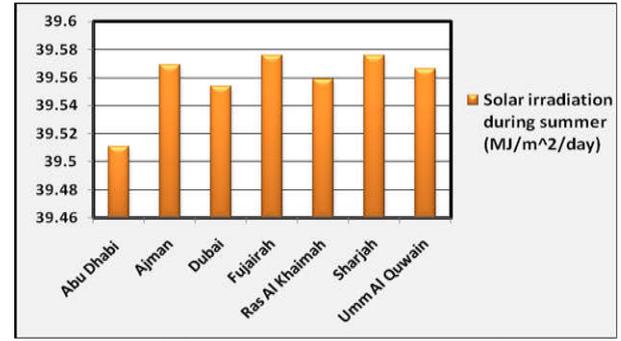

Fig. 3. Solar Irradiation during summer in UAE

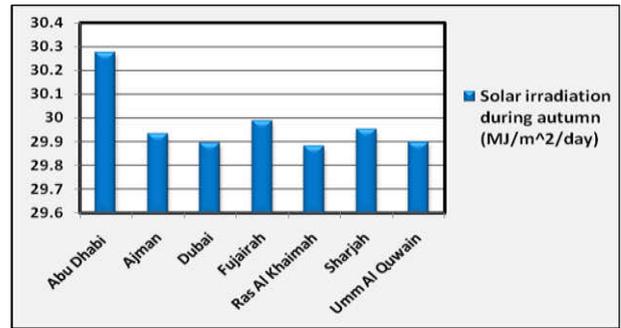

Fig. 4. Solar Irradiation during autumn in UAE

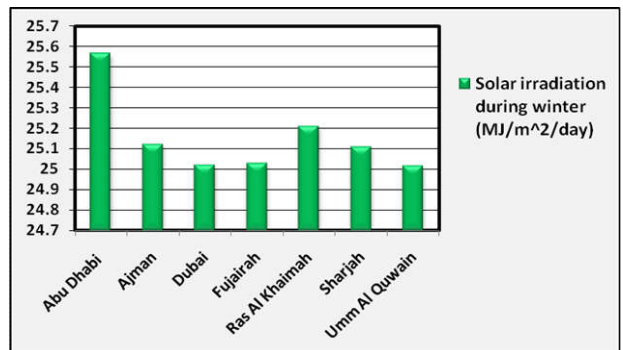

Fig. 5. Solar Irradiation during winter in UAE

After combining the four figures, we have found that the average solar irradiation of Abu Dhabi, Ajman, Dubai, Fujairah, Ras Al Khaimah, Sharjah, and Umm Al Quwain are 33.108MJ/m$^2$/day, 32.895MJ/m$^2$/day, 32.865MJ/m$^2$/day, 32.895MJ/m$^2$/day, 32.894MJ/m$^2$/day, 32.909MJ/m$^2$/day, and 32.858MJ/m$^2$/day respectively. Comparison between the maximum and minimum solar irradiation all over the country shows that Abu Dhabi and Sharjah have more solar irradiation than other emirates. Parallelly, the variation of solar irradiation in Dubai and Umm Al Quwain is less throughout the year than the other emirates. Comparably, less change in solar radiation helps to design a stand-alone PV module with less battery storage capacity and therefore reduces the cost of the module. During the summer (June-August) solar radiation values are at its highest and lowest during winter (December-February). The maximum radiation values for Abu Dhabi, Ajman, Dubai, Fujairah, Ras Al Khaimah, Sharjah, and Umm Al Quwain are 39.511MJ/m$^2$/day, 39.569MJ/m$^2$/day, 39.554MJ/m$^2$/day, 39.576MJ/m$^2$/day, 39.559MJ/m$^2$/day, 39.576MJ/m$^2$/day and 39.566MJ/m$^2$/day respectively and therefore should be consider as the input energy of the PV cell during summer (June-August). Similarly, the minimum radiation values for Abu Dhabi, Ajman, Dubai, Fujairah, Ras Al Khaimah, Sharjah and Umm Al Quwain are 25.566MJ/m$^2$/day, 25.124MJ/m$^2$/day, 25.021MJ/m$^2$/day, 25.029MJ/m$^2$/day, 25.211MJ/m$^2$/day, 25.112MJ/m$^2$/day and 25.020MJ/m$^2$/day respectively and therefore should be consider as the input energy of the PV cell during winter (December-February). Moreover, the radiation values for Abu Dhabi, Ajman, Dubai, Fujairah, Ras Al Khaimah, Sharjah and Umm Al Quwain that should be considered as the input energy of the PV cell during spring (March-May) are 37.080MJ/m$^2$/day, 36.956MJ/m$^2$/day, 36.992MJ/m$^2$/day, 36.989MJ/m$^2$/day, 36.925MJ/m$^2$/day, 36.995MJ/m$^2$/day and 36.945MJ/m$^2$/day respectively. And the radiation values for Abu Dhabi, Ajman, Dubai, Fujairah, Ras Al Khaimah, Sharjah and Umm Al Quwain that should be considered as the input energy of the PV cell during autumn (September-November) are 30.276MJ/m$^2$/day, 29.932MJ/m$^2$/day, 29.893MJ/m$^2$/day, 29.987MJ/m$^2$/day, 29.880MJ/m$^2$/day, 29.951MJ/m$^2$/day and 29.899MJ/m$^2$/day respectively.

## V. SIMULATION RESULTS OF PHOTOVOLTAIC CELL IN MATLAB

In this section, the I-V and P-V characteristics curves are obtained from the simulation for the MSX 120 PV cell at different solar irradiance and temperature. Here, for our research, we observe the performance parameters of MSX120 in seven different emirates in four different seasons: Spring (March-May), summer (June-August), autumn (September-November) and winter (December-February).

Since UAE is comparatively a small country compared to the other Gulf countries, all the seven emirates are located close to each other in terms of distance. Therefore, there is only a slight deviation in the solar irradiation and temperature between its seven emirates. The average temperature during spring in Abu Dhabi is 27.8°C and from the data in figure 2, the average solar radiation is 429watt/m$^2$. Again, the average temperature during spring in Ajman is 27.4°C and the corresponding solar radiation is 427watt/m$^2$. The average temperature and average solar radiation in Dubai during spring are 27.4°C and 428watt/m$^2$ respectively. The average temperature and average solar radiation in Fujairah during spring are 29.6°C and 428watt/m$^2$ respectively. Similarly, the average temperature and average solar radiation in Ras Al Khaimah, Sharjah and Umm Al Quwain during spring are 427watt/m$^2$ and 27.7°C; 428watt/m$^2$ and 29.3°C; 427watt/m$^2$ and 27.7°C respectively. Now again it has been found that the average temperature during summer in Abu Dhabi is 36.2°C and from the data in figure 3, the average solar radiation is 457watt/m$^2$. Similarly, the average temperature and the average solar radiation in Ajman, Dubai, Fujairah, Ras Al Khaimah, Sharjah and Umm Al Quwain during summer are: 35.7°C and 457watt/m$^2$; 35.7°C and 458watt/m$^2$; 37.4°C and 458watt/m$^2$; 35.4°C and 458watt/m$^2$; 38.1°C and 458watt/m$^2$; 35.4°C and 457watt/m$^2$ respectively.

Again, during autumn, it has been found that the average temperature in Abu Dhabi is 31.3°C and from the data in figure 4, the average solar radiation is 357watt/m$^2$. Similarly, for the same season the average temperature and the average solar radiation in Ajman, Dubai, Fujairah, Ras Al Khaimah, Sharjah and Umm Al Quwain are: 31.3°C and 346watt/m$^2$; 31.4°C and 345watt/m$^2$; 31.5°C and 347watt/m$^2$; 31.3°C and 345watt/m$^2$; 31.7°C and 347watt/m$^2$; 31.3°C and 346watt/m$^2$ respectively. Finally, the average temperature during winter in Abu Dhabi is 21.2°C and from the data in figure 5, the average solar radiation is 301watt/m$^2$. Similarly, the average temperature and the average solar radiation during winter in Ajman, Dubai, Fujairah, Ras Al Khaimah, Sharjah and Umm Al Quwain are: 21.2°C and 291watt/m$^2$; 22°C and 289watt/m$^2$; 21.8°C and 289watt/m$^2$; 22.8°C and 292watt/m$^2$; 21.4°C and 291watt/m$^2$; 22.8°C and 289watt/m$^2$ respectively.

Figure 6 and figure 7 illustrate the I-V and P-V characteristics curve of MSX120 influenced by the solar irradiation and temperature of Abu Dhabi. Figure 8 and figure 14, figure 9 and figure 15, figure 10 and figure 16, figure 11 and figure 17, figure 12 and figure 18, figure 13 and figure 19 illustrate the corresponding similar features of Abu Dhabi for Ajman, Dubai, Fujairah, Ras Al Khaimah, Sharjah and Umm Al Quwain respectively.

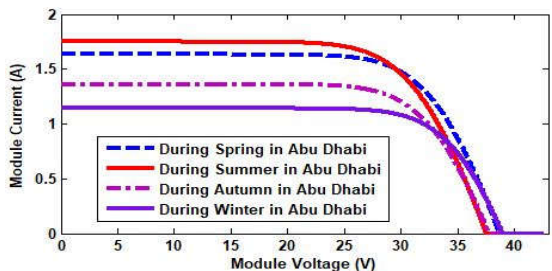

Fig. 6. Simulated I-V curve of MSX120 for Abu Dhabi

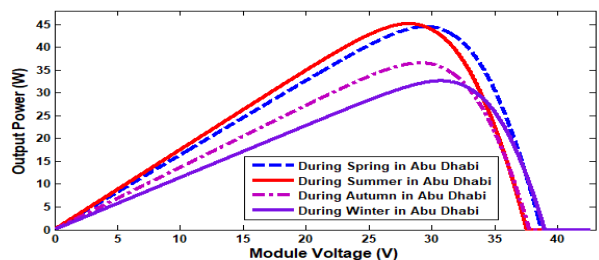

Fig. 7. Simulated P-V curve of MSX120 for Abu Dhabi

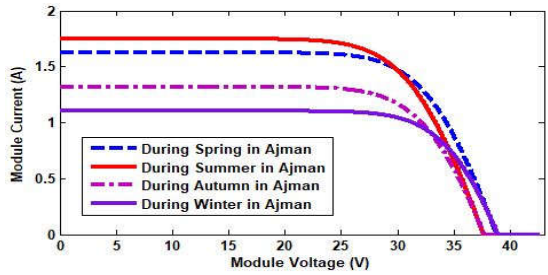
Fig. 8. Simulated I-V curve of MSX120 for Ajman

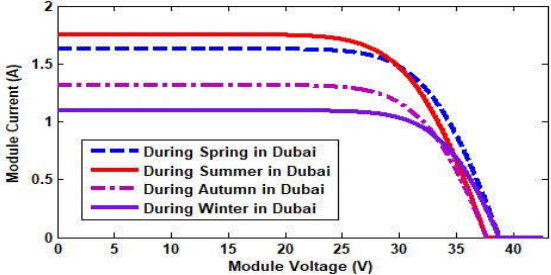
Fig. 9. Simulated I-V curve of MSX120 for Dubai

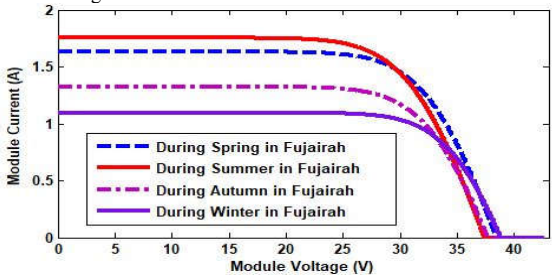
Fig. 10. Simulated I-V curve of MSX120 for Fujairah

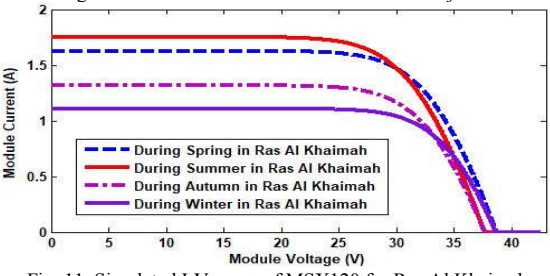
Fig. 11. Simulated I-V curve of MSX120 for Ras Al Khaimah

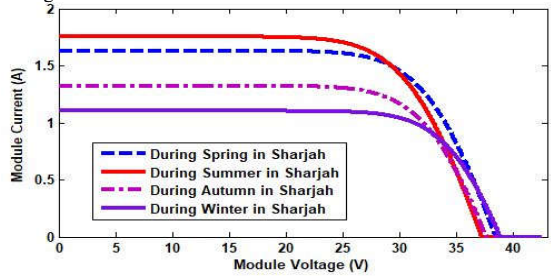
Fig. 12. Simulated I-V curve of MSX120 for Sharjah

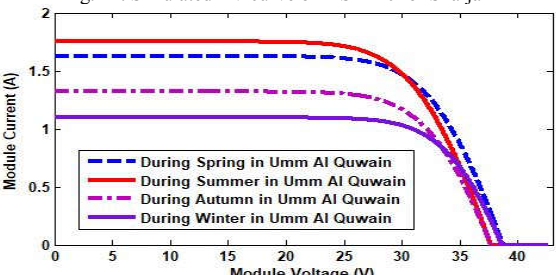
Fig. 13. Simulated I-V curve of MSX120 for Umm Al Quwain

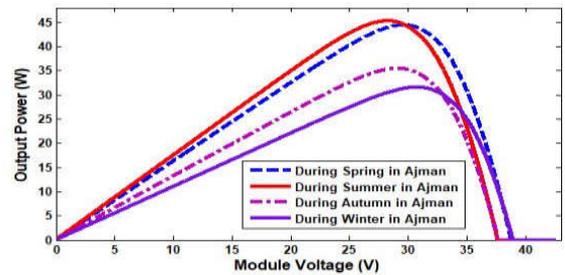
Fig. 14. Simulated P-V curve of MSX120 for Ajman

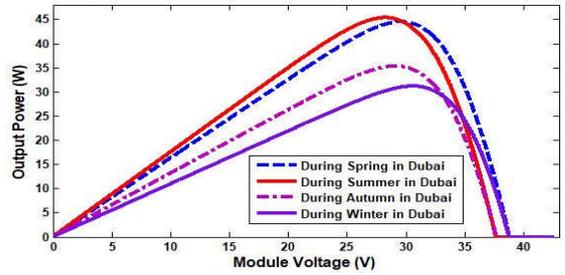
Fig. 15. Simulated P-V curve of MSX120 for Dubai

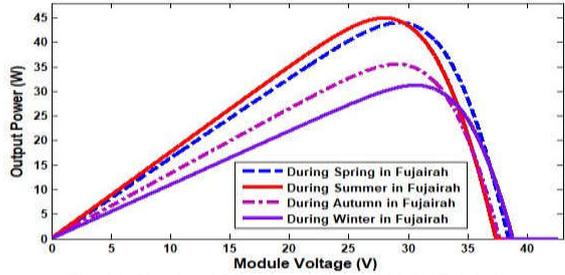
Fig. 16. Simulated P-V curve of MSX120 for Fujairah

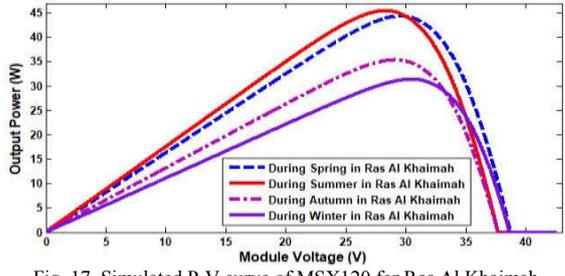
Fig. 17. Simulated P-V curve of MSX120 for Ras Al Khaimah

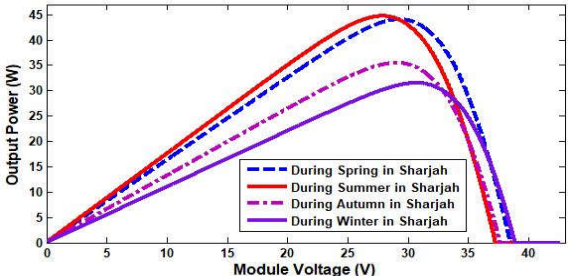
Fig. 18. Simulated P-V curve of MSX120 for Sharjah

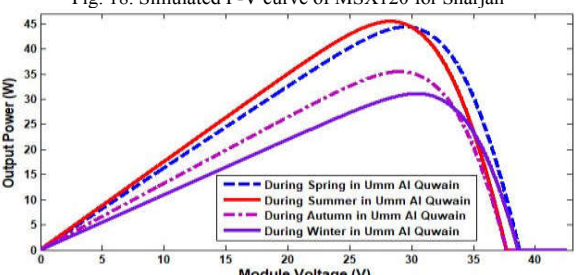
Fig. 19. Simulated P-V curve of MSX120 for Umm Al Quwain

## VI. PERFORMANCE EVALUATION ON THE SYSTEM

The performance of the PV module varies mainly due to environmental parameters like solar radiation, wind speed, humidity, and temperature. In this section, the experimental performance analysis of a PV system is presented. Here we have considered the average solar radiation for a day and night. Solar radiation is higher during day time therefore if we consider only day time then the performance of the PV cell will increase hence the average solar radiation value increases. From our above experimental results, we analyze the solar insolation and temperature data for seven emirates of UAE during spring, summer, autumn, and winter. Considering Abu Dhabi and Sharjah as the reference city, where the maximum power output of MSX120 during summer is found when the cell output voltage is around 28V and the corresponding maximum output power is about 45watt. Ajman, Fujairah and Ras Al Khaimah have lower maximum power output about 44 watts. And the lowest average power is found in Dubai and Umm Al Quwain about 43.5 watts during summer.

During spring and autumn, the output power is lower than the summer but the average output power during spring is a bit higher than the output power during autumn and during winter it becomes the lowest. Therefore, during spring the maximum power of Abu Dhabi and Sharjah is about 43watt and the maximum power of Ajman, Fujairah, and Ras Al Khaimah are about 42.9watt while the maximum power of Dubai and Umm Al Quwain are about 42.5watt during spring. During autumn the maximum power of Abu Dhabi and Sharjah is about 35watt and the maximum power of Ajman, Fujairah, and Ras Al Khaimah are about 34.2watt while the maximum power of Dubai and Umm Al Quwain is about 33.5watt during autumn. Again, during winter the maximum power of Abu Dhabi and Sharjah is about 30watt and the maximum power of Ajman, Fujairah, and Ras Al Khaimah are about 29watt while the maximum power of Dubai and Umm Al Quwain is about 28.9watt during winter.

From this section, it is observed that the variation of maximum power output in Dubai and Umm Al Quwain is less throughout the year which is very effective to design a PV cell with good battery storage capacity. But the maximum power output is obtained in Abu Dhabi and Sharjah during June-August. Therefore, if we can maintain good energy storage capacity of PV modules then Abu Dhabi and Sharjah are the recommended places to implement a number of solar PV modules as well as solar power plants.

## VII. CONCLUSION

In this paper, a double diode PV system with MSX120, 120W PV module is presented. I-V and P-V characteristics curves are obtained for the double diode PV module for different environmental and electrical parameters by Matlab simulation. Moreover, the maximum and minimum power output is obtained from the P-V characteristics for different places of UAE based on the variation of temperature and solar insolation. The maximum power point for each season and place are found when the voltage is about 28V. It is noticed that the output power changes in accordance with the solar irradiation and temperature, therefore, to get the maximum power during the daytime, Maximum Power Point Tracking should be installed. From the experimental data and the simulated result, it is observed that the variation of power and current is comparatively less in Dubai and Umm Al Quwain which is very useful to implement the large scale PV system to get the efficient power during the whole year. Thus, this paper provides a gateway to install a complete photovoltaic system in UAE. As UAE has many deserts located in different parts of the country, it is preferable to install a grid-connected PV system on desert areas which will provide clean energy as well as will reduce the power crisis in UAE.